\documentclass[12pt,a4paper]{article}
\usepackage{amssymb}
\usepackage{amsmath}
\usepackage{epsfig,graphics}
\def \dangle {\rangle \langle}
\def \be {{\begin{equation}}}
\def \ee {{\end{equation}}}
\def \bea {\begin{eqnarray}}
\def \eea {\end{eqnarray}}

\begin{document}
\title{Pseudo Memory Effects, Majorization and Entropy in Quantum Random Walks }
\author {Anthony J. Bracken$^{*}$, Demosthenes Ellinas$^{\dagger }$ and Ioannis
Tsohanjis$^{\ddagger}$\\
$^{*}$ University of Queensland,\\
Centre for Mathematical Physics \& Department of
Mathematics,\\Brisbane 4072 Australia\\ %
\smallskip  $^{\dagger \ddagger }$
Technical
University of Crete,  \\Divisions of
Mathematics$^{\dagger }$ and Physics$^{\ddagger }$  \\GR-731 00 Chania Crete Greece }
\maketitle

\begin{abstract}
A quantum random walk on the integers
exhibits pseudo memory effects, in that its probability distribution  after $N$ steps
is determined by
reshuffling the first $N$ distributions that arise in a classical random
walk  with the same initial distribution.
In a classical walk, entropy increase can be regarded as a consequence of
the majorization ordering of successive distributions.
The Lorenz curves of successive distributions for a symmetric quantum walk
reveal
no majorization ordering in general.  Nevertheless, entropy can increase,
and computer experiments show that it does so on average.
Varying the stages at which the quantum coin system is traced out
leads to new quantum walks, including a symmetric walk
for which majorization ordering is valid but the spreading
rate exceeds that of the usual symmetric quantum walk.
\end{abstract}

During the stochastic evolution of a classical random walk (CRW),
correlations are established among the states of its two constituent parts,
a \textit{coin} and a \textit{walker}. Because of the widespread use of the CRW
in applications involving
classical computer simulations \cite{hughes}, recent interest in quantum computation \cite
{nielsenchuang} has focussed attention on the notion of a quantum random
walk (QRW). The key idea is to replace the classical correlations between
coin and walker states in a CRW by the emblematic notion of quantum
correlation, \textit{i.e.}  entanglement \cite{nielsenchuang} of the states of
suitable quantum analogues of the coin and walker. Of main interest
in QRW studies has been the effect of entanglement on various asymptotics,
on spreading properties, and on hitting and mixing times. From the earliest
formulations of QRWs \cite{aharonov}\cite{mayer}, to recent studies on
general graphs \cite{ambainis}, on the line \cite{daharonov} \textit{etc.},
it has emerged that a number of surprising features distinguish quantum from
classical walks, such as their non-Gaussian asymptotics, a quadratic speed
up in spreading rate on the line \cite{nayak}, an exponentially faster
hitting time in hypercubes \cite{moore}\cite{kempe}, and an exponentially
faster penetration time of decision trees \cite{childs1}\cite{childs2}.
These findings have recently prompted proposals for physical implementation
of such processes in experiments, {\it e.g.} in ion traps \cite{travaglione},
optical lattices \cite{dur}, or in cavity QED \cite{sanders}. A
review providing a comprehensive introduction and other references has recently appeared
\cite{kempeReview}.

There are two different ways to consider a QRW as a process.
In the first,
a fixed number $N$ of
applications of a unitary quantum evolution to a combined coin-walker
system is considered, producing a
highly entangled state, and after the last step, the coin degrees of
freedom are traced out
to determine a reduced density matrix and subsequently
$P_Q^{(N)}$,
the associated probability distribution (pd)
of location probabilities
for the walker.
In the second, extended interpretation,
the QRW is considered as the process that produces
{\it successively} in this way, the pds $P_Q^{(N)}$ for $N=0,\,1,\,2,\,3,\,\dots$.
The purpose of the present note is to
indicate some further remarkable properties that differentiate
a QRW, interpreted in this extended way,  from a CRW.

We construct the QRW on $H=H_{c}\otimes H_{w}$,
where the coin space is $H_{c}$ $=l_{2}(\{0,1\})$ and the walker space is $H_{w}=l_{2}(Z)$,
and we consider the unitary evolution operator $%
V=P_{+}U\otimes E_{+}+P_{-}U\otimes E_{-}$ acting on $H$.  Here $U$ is $2\times 2$ unitary, and
acts on $H_c$ along with $P_{+}$
and $P_-$, which are orthogonal projectors onto the coin states $|0\rangle = (1\,\,0)^T$ and
$|1\rangle = (0\,\,1)^T$ respectively.  The step operators
$E_{+}$, $E_{-}$
act on $| m\rangle\in H_w$ for $m=0,\,\pm 1,\,\pm 2,\,\dots\,$
as $E_{\pm }| m\rangle= | m\pm 1\rangle,$ so that $E_{\pm }E_{\mp }=\mathbf{%
1}$.  Together with the \textit{distance operator}  $L$ acting as $%
L| m\rangle=m| m\rangle,$
they satisfy the commutation relations $[L,E_{\pm }]=\pm E_{\pm },$ $[E_{+},E_{-}]=0.$
With
$\rho^{(0)} _{c}= | \varphi \dangle \varphi | $ chosen as the initial density matrix of
the coin and $\rho _{w}^{(0)}=| 0\dangle 0| $ the initial density matrix of
the walker, the total initial density matrix $\rho ^{(0)}$ is
\begin{equation}
\rho ^{(0)}=\rho^{(0)} _{c}\otimes \rho _{w}^{(0)}=| \varphi \dangle \varphi |
\otimes | 0\dangle 0|\, .  \label{rho0}
\end{equation}
After $N$ successive applications of the unitary evolution defined by $V$,
the coin degrees of freedom are traced out, so that
$\rho _{w}^{(0)}$ evolves to $\rho _{w}^{(N)}$,
given by the action of a trace preserving completely positive
(CP) map $\varepsilon _{V^{N}}$ \cite{kraus}
\begin{eqnarray}
\rho _{w}^{(N)}&=&\varepsilon _{V^{N}}(\rho _{w}^{(0)})\equiv \
Tr_{c}[V^{N}(\rho^{(0)} _{c}\otimes \rho _{w}^{(0)})(V^{\dagger
})^{N}]
\nonumber
\\
&=&\sum_{k=0,1}A_{k}^{(N)}\rho _{w}^{(0)}A_{k}^{(N)\dagger } \,.
\label{rsn}
\end{eqnarray}
Here the Kraus operators are given for $k=0,\,1$ by
$A_k^{(N)}=\langle k| V^N | \varphi \rangle$,
and satisfy
\begin{equation}
\sum_{k=0,1}A_k^{(N)%
\dagger }A_k^{(N)}=\mathbf{1}\,.  \label{kr}
\end{equation}
The diagonal element
$\langle k |\rho _{w}^{(N)}|k\rangle$
of the walker density matrix,
for $k\in\{-N,\,-N+2,\,\dots \,N\}$,
is the
probability
$P_Q^{(N)}(k)$
of occupation
of the site $k$
by the walker after the $N$th step.
It follows from (\ref{rho0}) that $P_Q^{(0)}(k)=\delta_{k0}$.

With $V$ defined as above we have
$ V^{N}=\left(
\begin{array}{cc}
\alpha ^{(N)} & \beta ^{(N)} \\
\gamma ^{(N)} & \delta ^{(N)}
\end{array}
\right)
\label{Vform}$,
for suitable operators $\alpha ^{(N)}$ {\it etc.} acting  on $H_{w}$.
For the general  choice $| \varphi \rangle=c| 0\rangle +d| 1\rangle,$ with $c$, $d$
complex,
we have that
$A_{0}^{(N)}=c\alpha ^{(N)}+d\beta
^{(N)},\;A_{1}^{(N)}=c\gamma ^{(N)}+d\delta ^{(N)}.$
We choose
$U=U(p)=\left(
\begin{array}{cc}
\sqrt{p} & \sqrt{1-p} \\
\sqrt{1-p} & -\sqrt{p}
\end{array}
\right)$, with $0\leq \;p\leq 1,$
without any significant loss of generality \cite{ambainis}.
The choice $c=1/\sqrt{2}$, $d=i/\sqrt{2}$
and $p=1/2$
is known
\cite{travaglione,ambainis}
to result in
$P_Q^{(N)}(-k)=P_Q^{(N)}(k)$ for all $k$,
but a  QRW
is not symmetric in general.
In what follows, we utilize the elementwise or Hadamard product $%
\Delta =A\circ B$, defined between matrices $A$,$B$ of the same size by $%
(A\circ B)_{ij}=A_{ij}B_{ij}$, and we call {\it doubly stochastic} \cite{bhatia}
a square matrix $\Delta$
(of
finite size, or with columns and rows of finite support, as is the case here),
with nonnegative elements, which has
unit column and row sums.

Our first result is summarized as:

\textit{Proposition 1.} \textit{\ There exists a doubly stochastic
matrix}
$\Delta _{Q}^{(N)}=A_{0}^{(N)}\circ
\overline{A}_{0}^{(N)}+A_{1}^{(N)}\circ \overline{A}_{1}^{(N)},$ \textit{\
that connects the initial pd } $P_Q^{(0)}$
\textit{\ with
the pd  } $P_Q^{(N)}$
\textit{\ of the } $N$\textit{th step.}
\textit{\ These matrices
satisfy the inhomogeneous recurrence
relation}
\begin{eqnarray}
\Delta _{Q}^{(N+1)}& =&A_{0}^{(N+1)}\circ \overline{A}%
_{0}^{(N+1)}+A_{1}^{(N+1)}\circ \overline{A}_{1}^{(N+1)}  \notag \\
& =&((1-p)E_{+}+pE_{-}))\Delta _{q}^{(N)}  \notag \\
 +\,(E_{+}-E_{-})&&\!\!\!\!\!\!\!\!\!\![\sqrt{p(1-p)}(A_{0}^{(N)}\circ \overline{A}%
_{1}^{(N)}+A_{1}^{(N)}\circ \overline{A}_{0}^{(N)})\notag \\
&+&(2p-1)A_{0}^{(N)}\circ
\overline{A}_{0}^{(N)}].  \label{Deltaform}
\end{eqnarray}
\textit{\ By means of this relation,} $%
P_{Q}^{(N)}$ \textit{ is related to the
classical pds }
$\{P_{C}^{(N)},\;P_{C}^{(N-1)},...,\;P_{C}^{(1)},\;P_{C}^{(0)}
=P_{Q}^{(0)}
\}$,
\textit {\ arising from the first} $N$ \textit{\ steps of a CRW,
via the map } $\delta ^{(N)}$ \textit{\ given by }
\begin{eqnarray}
P_{Q}^{(N)} &= &\delta
^{(N)}(P_{C}^{(N)},...,P_{C}^{(0)}
)
\nonumber\\
&=&P_{C}^{(N)}+\omega
^{(1)}P_{C}^{(N-1)}+\omega ^{(2)}P_{C}^{(N-2)}
\notag
\\
&&+...+\omega ^{(N-1)}P_{C}^{(1)}+\omega ^{(N)}P_{C}^{(0)},
\label{theorem1}
\end{eqnarray}
\textit{where we have introduced the reshuffling matrices }
$\omega
^{(i+1)}=(E_{+}-E_{-})M^{(i)},\quad i>0$
\textit{with }
$M^{(i)}=\sqrt{p(1-p)}%
(A_{0}^{(i)}\circ \overline{A}_{1}^{(i)}+A_{1}^{(i)}\circ \overline{A}%
_{0}^{(i)})+(2p-1)A_{0}^{(i)}\circ \overline{A}_{0}^{(i)}$
\textit{\ and with}
$%
M^{(0)}=0$, \textit{\ } $\omega ^{(1)}=0$.

The proof is straightforward.

\textit{Comments}:

1) Since the $A_{0,1}^{(N)}$
are polynomials of degree $N$ in the commuting step operators $E_{\pm }$, they are
normal operators {\it i.e.} $A_{k}^{(N)}$ $%
A_{k}^{(N)\dagger }$ $=$ $A_{k}^{(N)\dagger }$ $A_{k}^{(N)}$ for $k=0,1.$
This means that in addition to (\ref{kr}), there is a similar
relation with $A_k^{(N)}$, $A_k^{(N)\dagger}$ interchanged.
Together,  these two
relations lead to the double stochasticity of $\Delta _{Q}^{(N)}$
\cite{chefles}.  This simple method of
constructing doubly stochastic matrices by convex sums of Hadamard products
of Kraus generators of CP maps, being normal operators, together with the question of
the ensuing entropy increase, to be discussed shortly, is a nontrivial
extension of Uhlmann's theory which addresses those questions for unitary
CP maps only (\textit{cf.} \cite{uhlmann} and references therein).

2) The  general
form  of (\ref{Deltaform}) is $\Delta _{Q}^{(N+1)}=\Delta
_{C}\Delta _{Q}^{(N)}+(E_{+}-E_{-})M^{(N)}$, and we see that the final $N$%
-dependent inhomogeneous term, which must have zero column and row sums, distinguishes
a QRW from a CRW and moreover carries the burden of possible breaking of the
majorization ordering, as will be seen shortly.

3) Proposition 1 shows that the effect of
tracing out the coin system after $N$ applications of $V$
to $\rho^{(0)}$,
results in a kind of pseudo memory effect (or
pseudo
nonMarkovian effect),  in that determination of the quantum occupation probabilities
at step $N$
involves the occupation probabilities of an $N$-step CRW,
reshuffled from step to step as in (\ref{theorem1}).
This suggests a modified QRW where the coin system is traced
out after every $m$ steps, for some fixed $m$, rather than after $1$ or $2$ or $\dots$ $N$ steps as
in the QRW as considered to date.
The case $m=1$ defines a scheme
that \textit{promptly traces} the coin system after each $V$
action. With $U(p)$ as above and $\rho^{(0)}$ as in (\ref{rho0}),
this yields the occupation probabilities of a $(p,1-p)$ biased
CRW. Indeed if $\rho _{w}^{(N+1)}=(\varepsilon _{V})^{N+1}(\rho
_{w}^{(0)})=\varepsilon _{V}(\rho _{w}^{(N)})\equiv \ Tr_{c}[V(\rho
_{c}^{(0)}\otimes \rho _{w}^{(N)})V^{\dagger }],$ then we have at each step,
along the diagonal of the reduced density matrix,  the probabilities
of the corresponding row of the classical Pascal triangle {\it i.e.} $\langle k|\rho
_{w}^{(N)}|k\rangle =(P_{Q}^{(N)})(k)=(P_{C}^{(N)})(k)$. In this case no
memory effects are present and $P_{Q}^{(N+1)}=\Delta _{C}P_{Q}^{(N)},$ where
$\Delta _{C}=(1-p)E_{+}+pE_{-}$ is a doubly stochastic matrix
that repeatedly mixes the evolving probability distribution and
extends its support by one unit to the left and right at each time step of
the walk.

4)
Proposition 2 below is devoted to the case
$m=2$, where a pseudo memory effect is also exhibited, since here also the
determination of the quantum pd at step $N$ involves
classical occupation probabilities from the first $N$ steps. However now the
reshuffling matrix $\Phi $ is fixed (as in (\ref{phi}) below).

\textit{Proposition 2.\ There exists a doubly stochastic matrix
} $\Delta _{Q}=B_{0}\circ \overline{B}_{0}+B_{1}\circ
\overline{B}_{1}$\textit{\ that connects the }$N$th\textit{\ step
pd }$P_{Q}^{(N)}$\textit{\ identified with the
diagonal elements of }$\rho _{w}^{(N)},$\textit{\ with }$P_{Q}^{(N+1)}$%
\textit{\ at the }$(N+1)$ $th$\textit{\ step, identified with the diagonal
elements of }$\rho _{w}^{(N+1)}=\varepsilon _{V^{2}}(\rho _{w}^{(N)})$.%
 \textit{Here the Kraus generators are}
\begin{align*}
B_0& =(pc+\sqrt{p(1-p)}d)E_{+}^2\!+\!((1-p)c-\sqrt{p(1-p)}d)\mathbf{1},\; \\
B_1& =(pd-\sqrt{p(1-p)}c)E_{-}^2\!+\!((1-p)d+\sqrt{p(1-p)}c)\mathbf{1}.
\end{align*}
\textit{\ The recurrence relation satisfied by }$%
P_{Q}^{(N)}$, \textit{\ and its solution, are given by
}
\begin{align}
&P_{Q}^{(N)} =\Delta _{Q}P_{Q}^{(N-1)}=(\Delta _{C}+\Phi )P_{Q}^{(N-1)},%
\text{ and}  \notag \\
&P_{Q}^{(N)} =\Delta _{Q}^{N}P_{Q}^{(0)}=(\Delta _{C}+\Phi
)^{N}P_{Q}^{(0)}\notag \\
&=\sum_{k=0}^{N}\left(
\begin{array}{c}
N \\
k
\end{array}
\right) \Phi ^{N-k}\Delta _{C}^{k}P_{Q}^{(0)}=\sum_{k=0}^{N}\left(
\begin{array}{c}
N \\
k
\end{array}
\right) \Phi ^{N-k}P_{C}^{(k)},
\notag
\end{align}
\textit{where }$\Delta _{Q}=\Delta _{C}+\Phi ,$\textit{\ and where the
matrix }$\Phi $\textit{\ with  null column and row sums
is given explicitly by }
\begin{align}
\Phi & =B_{0}\circ \overline{B}_{0}+B_{1}\circ \overline{B}_{1}-\Delta _{c}
\notag \\
& =|pc+\sqrt{p(1-p)}d|^{2}E_{+}^{2}+|\sqrt{p(1-p)}c-pd|^{2}E_{-}^{2}
\notag \\
& +(|(1-p)c-\sqrt{p(1-p)}d|^{2}+|(1-p)d \notag \\
&+\sqrt{p(1-p)}c|^{2})\mathbf{1}%
-(1-p)E_{+}-pE_{-}\text{.}  \label{phi}
\end{align}
Once it is seen that in this case the analogue of (\ref{rsn}) reads
$\rho _w^{(N+1)}=\varepsilon _{V^2}(\rho _w^{(N)})\equiv \ Tr_c[V^{2N}(\rho
_c^{(0)}\otimes \rho _w^{(0)})(V^{2\dagger })^N]=\sum_{k=0,1}B_k\rho
_w^{(0)}B_k^{\dagger }$, the rest of the proof
is similar to that for the Proposition 1.
Generalization to the case $m>2$ is straightforward.

\begin{figure}
\centerline{\psfig{figure=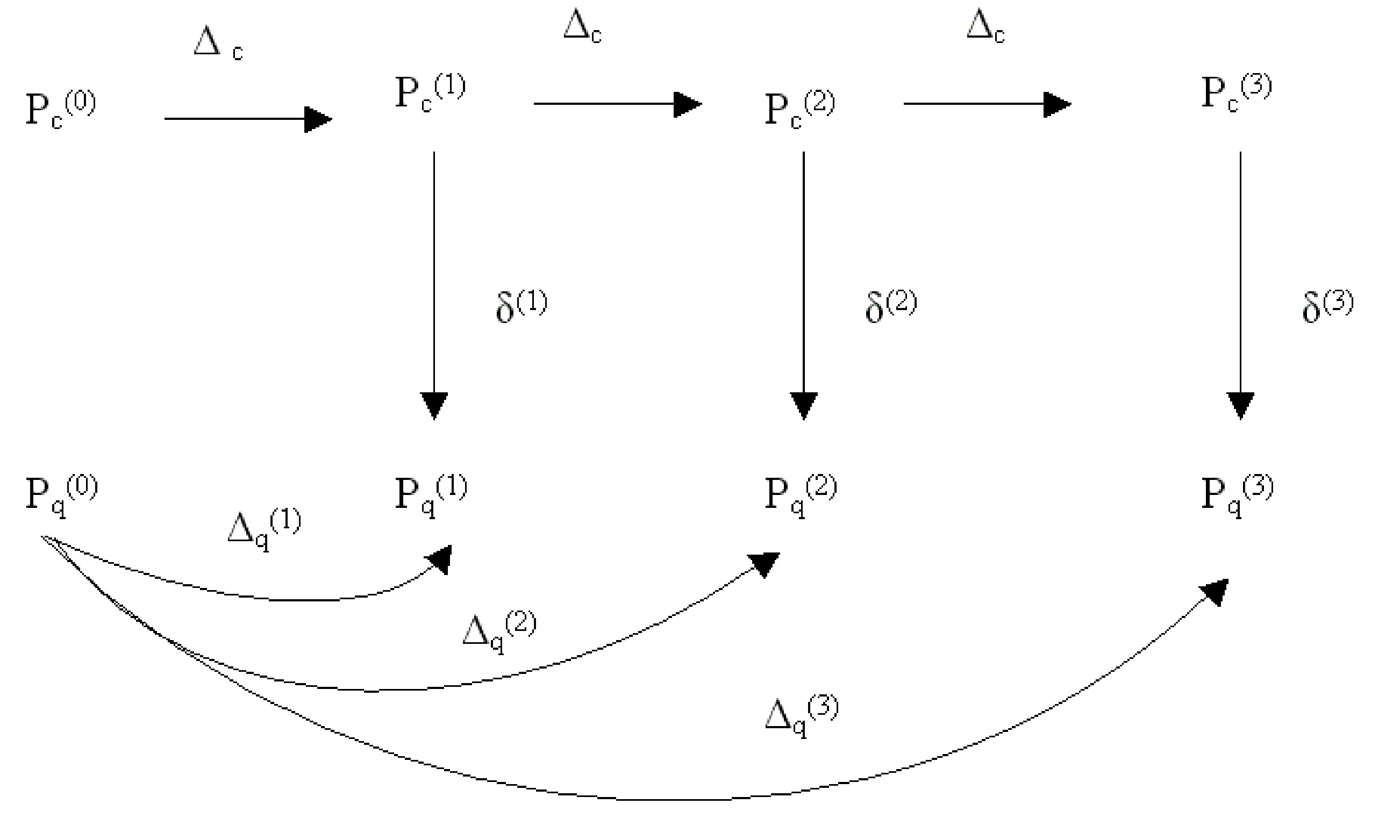,height=85mm,width=155mm}}
\rotatebox{90}{}
\caption
{Successive distributions of a CRW (upper
horizontal line) obtained by action of $\Delta _c$, and of a QRW as in  Proposition 1
(lower horizontal line), obtained by the action of $\Delta _q^{(N)}$. The
pseudo memory effect is shown by the vertical arrows $\delta ^{(N)}.$  }
\label{fig1}
\end{figure}

\begin{figure}
\centerline{\psfig{figure=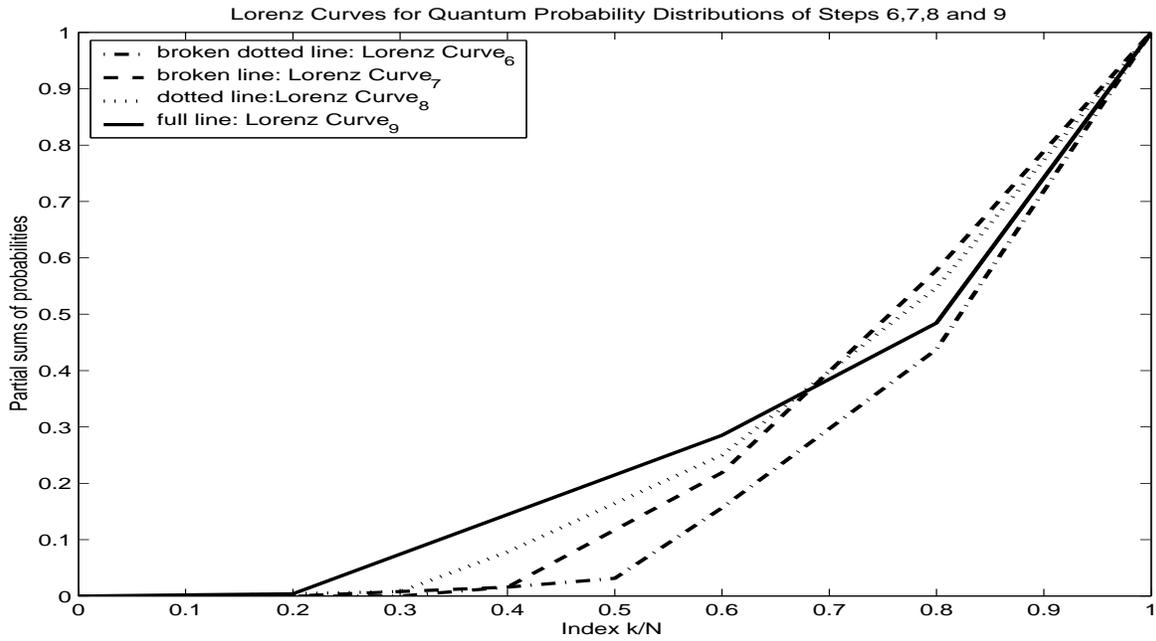,height=85mm,width=155mm}}
\caption{Lorenz curves for the distributions of steps 6,7,8 and 9 of the symmetric QRW
of Proposition 1.
}
\label{fig2}
\end{figure}

It is known that CRW pds
become more entropic as $N$ increases. This can be attributed to
the fact that they are ordered by
majorization \cite{marshalolkin}. Thus, for two consecutive classical pds $P_{C}^{(N)}$ and $%
P_{C}^{(N+1)},$ each with elements arranged in nondecreasing order, it is true that
$P_{C}^{(N)}\succ P_{C}^{(N+1)}$, and \textit{therefore} that
$S(P_{c}^{(N)})\leq
S(P_{c}^{(N+1)}),$ for the respective Shannon entropies,
defined as $S(P)=-\sum_k P(k)\log P(k)$. To facilitate a
comparison with the corresponding behaviour in the symmetric  QRW of Proposition 1, we set
the  classical pd $P_{C}^{(N)}$ in the  upper horizontal line in Fig. \ \ref{fig1}, and
the quantum pd $P_{Q}^{(N)}$
in the lower horizontal line.
We find the remarkable result that in certain cases, though QRW pds are
becoming more entropic in the course of time, namely $S(P_{Q}^{(N)})\leq
S(P_{Q}^{(N+1)})$, majorization breaks down \textit{i.e.} $P_{Q}^{(N)}\nsucc
P_{Q}^{(N+1)}$, even in the early stages.
This is illustrated in
Fig. \ \ref{fig2}, which refers to steps $N=6,\,7,\,8$ and $9$ of
the symmetric QRW.  Here the entropy is increasing, with
$S^{Q}(6)\approx 1.6551,\,S^{Q}(7)\approx 1.8138,\,S^{Q}(8)
\approx 1.8909,\,S^{Q}(9)\approx 1.9295$,
but majorizarion ordering is violated, as
can be seen at once from the corresponding
Lorenz curves. The  Lorenz curve of a pd $P$ whose elements $P(k)$
have been arranged in nonincreasing order
is the plot of the points $(n/N,\,
\gamma_{n}(P))$
for $n=0,\,1,\, \dots\, N$,
where $\gamma_{n}(P)=\sum_{k=1}^{n}P(k)$,
and $\gamma_{0}(P)=0,\, \gamma_{N}(P)=1$.
If $\,P\prec P'$ then the Lorenz curve of $P'$ always lies below that of $P$ and never crosses it.
Thus \textit{crossing of Lorenz curves implies majorization breakdown and vice versa}.
Such a breakdown of majorization takes
place in the symmetric QRW as is seen in Fig. 2, while entropy increases as seen in Fig. 3.
Remarkably, this is not the case for the QRW of Proposition 2, since in this
case the mixing matrix from step to step remains $N$-independent, fixed and
doubly stochastic, guaranteeing majorization ordering \cite{marshalolkin,nielsen}.

\begin{figure}
\centerline{\psfig{figure=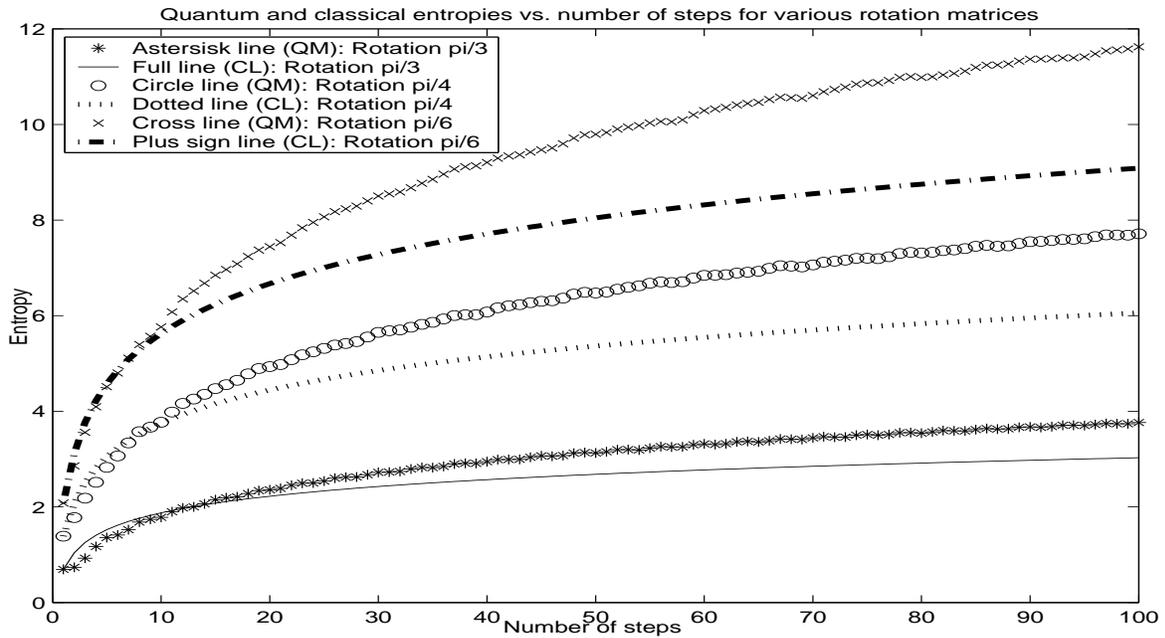,height=85mm,width=155mm}}
\caption{Quantum and classical entropies
{\em v.} number of steps. The symmetric QRW is as in  Proposition 1 in each case.}
\label{fig3}
\end{figure}

Fig. \ \ref{fig3} illustrates three notable features of
entropy dynamics common to three symmetric QRWs as in Proposition 1, with  $U(p)$
having $p=1/3=\cos^2(\pi/6),\, p=1/2=\cos^2(\pi/4)$ and $p=3/4=\cos^2(\pi/3)$, respectively:
firstly, \textit{an increase of entropy on average, e.g.} in the case $p=1/4$,
the sequence of steps $%
45,\,51,\,57,\,63,\,\dots $
has monotonically increasing entropy values;
secondly,
clusters of steps with \textit{decreasing entropy, e.g. } $S(P_Q^{(49)})\approx 3.3498,%
\;S(P_Q^{(50)})\approx 3.3467,\;S(P_Q^{(51)}\approx 3.3408$;
and thirdly,
\textit{a larger rate of increase} of quantum entropy
\textit{cf.}  classical entropy on average.

Finally, we indicate that a symmetric QRW with a constantly delayed tracing
scheme, as in Proposition 2,  shows an even greater rate of
spreading than the usual symmetric QRW,
which in turn is
known to spread quadratically faster than  a CRW
\cite{nayak,ambainis}.

Define the $m$th order statistical moment of the  the distance
operator $L$ at step $N$ by $\langle L^{m}\rangle _{N}\equiv
Tr(\rho _{w}^{(N)}L^{m})$, where for the ``classical" case corresponding to the ``promptly traced" QRW,
$\rho
_{w}^{(N)}=(\varepsilon _{V})^{N}(\rho _{w}^{(0)})$, and for the cases of
quantum walks described in Propositions 1 and 2, $\rho _{w}^{(N)}=\varepsilon
_{V^{N}}(\rho _{w}^{(0)})$ and $\rho _{w}^{(N)}=(\varepsilon _{V^{2}})^N(\rho
_{w}^{(0)})$, respectively.
Consider symmetric walks in each case, with $p=1/2$.
All first moments are zero in each case,
\textit{i.e.} $\langle L\rangle _{N}=0$, for all $%
N $, so that the standard deviation  at step $N$ is given by $\sigma
_{N}=\sqrt{\langle L^{2}\rangle _{N}}$.
For the CRW we have $\sigma _{N}^{C}=%
\sqrt{N}$ as is well known \cite{hughes}.
Our methods allow us to calculate $\sigma_N$ easily for
QRWs, for a given $N$.  For the first five steps of the
QRW of Proposition 1 we get
$\sigma _{1}^{Q1}=\sigma_{1}^C,\;\sigma _{2}^{Q1}=\sigma_2^C,
\;\sigma _{3}^{Q1}=\sigma_3^C,
\;\sigma _{4}^{Q1}=
(\sqrt{5}/2)\,\sigma_4^C,\;\sigma _{5}^{Q1}=\sqrt{8/5}\,\sigma_5^C$.
The enhanced rate of growth in the
quantum case, which is known
\cite{nayak,ambainis} to be given as $N\to\infty$ by
$\sigma_N^{(Q1)}\sim \sqrt{N (2-\sqrt{2})/2}\,\sigma_N^C$,
is soon clear.

For the first five steps of the
QRW of Proposition 2 we  have
$\sigma _{1}^{Q2}=\sigma_1^C,\;\sigma _{2}^{Q2}=\sqrt{5/2}\sigma_2^C,\;
\sigma _{3}^{Q2}=\sqrt{3}\sigma_3^C,\;
\sigma
_{4}^{Q2}=\sqrt{7/2}\sigma_4^C,\;
\sigma
_{5}^{Q2}=2\sigma_5^C$.

We see that the standard deviations for this `delayed tracing'
QRW  grow even
faster than those of the first type of QRW.
Since this second type of walk has constant Kraus generators, it may well be more easily
implemented experimentally than the first type
\cite
{travaglione,dur,sanders}.

\bigskip \bigskip

Acknowledgment: D. E. and I. T.  thank the Department of
Mathematics, University of Queensland for kind hospitality, and  I.T.
acknowledges an Ethel Raybould Fellowship from that Department.

\end{document}